\documentclass[letterpaper]{iopart}
\usepackage{amsfonts}
\usepackage{amsmath}
\usepackage{setspace}
\usepackage{graphicx}
\usepackage{psfrag}

         \usepackage{graphics}
\def\scrip{{\Im^+}}
\def\scrim{{\Im^-}}
\begin{document}
\title{A Dynamical Systems Approach to Schwarzschild Null Geodesics}


\date{March 1, 2011}

\author{Edward Belbruno${}^{1,2}$ and Frans Pretorius${}^3$}
\medskip
\address{\footnotesize
{${}^1$ Courant Institute of Mathematical Sciences, New York University}\\
{${}^2$ Princeton University}\\
{${}^3$ Department of Physics, Princeton University}
}

\medskip

\bigskip

\begin{abstract}
The null geodesics of a Schwarzschild black hole are studied from a dynamical
systems perspective. Written in terms of Kerr-Schild coordinates, 
the null geodesic equation takes on the simple form of a 
particle moving under the influence of a Newtonian central force with an 
inverse-cubic potential. We apply a McGehee transformation to these equations,
which clearly elucidates the full phase space of solutions. 
All the null geodesics belong to one of four families of invariant 
manifolds and their limiting cases,
further characterized by the angular momentum $L$ of the orbit:
for $|L|>|L_c|$, (1) the set that flow outward from the white hole, turn
around, then fall into the black hole, (2) the set that fall inward from past null 
infinity, turn around outside the black hole to continue to future null infinity, and
for $|L|<|L_c|$, (3) the set that flow outward from the white hole and continue
to future null infinity, (4) the set that flow inward from past null infinity
and into the black hole. The critical angular momentum $L_c$ corresponds
to the unstable circular orbit at $r=3M$, and the homoclinic orbits associated with it. 
There are two additional critical points of the flow at the singularity
at $r=0$. Though the solutions of geodesic motion and Hamiltonian
flow we describe here are well known, what we believe is a novel aspect of this work is
the mapping between the two equivalent descriptions, and the different insights each
approach can give to the problem. For example, the McGehee picture points to a particularly
interesting limiting case of the class (1) that move from the white to black hole: in the 
$L\rightarrow \infty$ limit, as described in Schwarzschild coordinates, these geodesics 
begin at $r=0$, flow along $t=\rm constant$ lines, turn around at $r=2M$,
then continue to $r=0$. During this motion they circle in azimuth exactly once, and complete
the journey in zero affine time.

\end{abstract}

\maketitle

\section{\bf{Introduction}}
\label{Section:1} 

\setlength\oddsidemargin{4pc}
\setlength\evensidemargin{4pc}

The Schwarzschild metric, describing a static, non-rotating black hole solution to the
Einstein field equations, was discovered within a year of Einstein's publication
of the theory of general relativity. Though it had to wait till the 1960's before
the full nature of the metric was truly uncovered, it has nevertheless
been well studied for nearly a century. One of the more important tools in this
regard is understanding the geodesic structure of the spacetime; in particular,
timelike and null geodesics characterize the paths that a freely moving test
particle $P$ can follow within the spacetime. The geodesic equations of motion
for $P$ are a set of second order, ordinary
differential equations, describing the evolution of the coordinates of $P$
as a function of an affine parameter. In Schwarzschild these equations are integrable,
and the solutions have been known for a long time (see any standard text on general
relativity). Nevertheless, the solutions exhibit a sufficiently rich
set of dynamics that new insights into them are still being garnered.
For example, in the past couple of decades dynamical systems methods have been
brought to bare on the geodesic equations, and perturbations thereof.
A couple of the interesting results have been a new taxonomy of 
orbits based on the subset of orbits that are periodic, emphasizing the importance
of the homoclinic orbits that asymptotically approach the unstable branch 
of circular orbits~\cite{Levin:2008mq,Levin:2008yp,PerezGiz:2008yq},
and that a generic perturbation of the geodesic flow possesses a chaotic invariant 
set~\cite{Moeckel,Vieira:1996zf,Letelier:1996he,Cornish_Frankel,Suzuki:1999si,deMoura:1999wf,Saa:1999je}.

In this paper we describe a new approach to understand the phase space
of geodesic orbits, by applying a transformation due to McGehee \cite{McGehee}.
This transformation is designed
to resolve the singularities that formally appear in the Newtonian equations
of motion when particles interacting through a central force collide.
The transformation ``blows up'' collision into an invariant manifold for the flow of 
the transformed differential equations. This allows the motion of $P$ to be studied near collision, 
uncovering interesting dynamics. 
To apply these methods, we map the geodesic equation, written in Cartesian-like 
Kerr-Schild~\cite{Kerr_Schild} 
(or ingoing Eddington-Finkelstein~\cite{Eddington,Finkelstein:1958zz})
coordinates, to a central force problem. This is trivial for the case where $P$ has zero mass 
(null geodesic), which we focus on in this paper, though the methods can be generalized. 

The rest of the paper is laid out as follows. In Section \ref{Section:1.5} we introduce the Schwarzschild metric
in the standard Schwarzschild coordinates,
give an overview of its Penrose diagram, and qualitatively discuss the null geodesics
of the geometry. This will provide a point of reference as we map the geometric
picture to a dynamical systems description within McGehee coordinates.
In Section~\ref{Section:2} the differential equations for the motion of $P$ are 
given, in Schwarzschild and Kerr-Schild coordinates. In Section
\ref{Section:3} we briefly summarize the results of the McGehee method for understanding the flow
of a general class of Hamiltonian systems corresponding to central force problems.
In Section \ref{Section:3.5} we restrict to the particular system that maps to the
null geodesic structure of Schwarzschild. We show that the phase
space flow can be subdivided into four families of invariant manifolds and their
limiting cases. There are four critical points of the flow---the black hole
and white hole singularities at $r=0$, and the 
two (with angular momentum $L=\pm L_c$) unstable circular orbits at
$r=3M$. Though much of the understanding of geodesic motion and the 
classical Hamiltonian system are individually  well known, we believe the novel aspect of this work is
the (sometimes non-trivial) mapping between the two equivalent descriptions,
and the insight one gives to the other. 
One particularly interesting example is the limiting case of geodesics that
in the dynamical systems description flow ``directly'' from the the white
to black hole: the collective set of these geodesics trace out the
interior region of the white/black hole, flow along Schwarzschild $t=\rm const.$
lines, and complete exactly one orbit $\Delta\Phi=2\pi$ along the journey
from white hole to black hole singularity. Also, the standard
affine parameter integrates to zero along these curves. To obtain
a finite affine length requires rescaling it 
by an angular momentum dependent quantity, which diverges in the limit.
Details of all this are discussed in
Section \ref{Section:3.5}.
Finally, we conclude in Section \ref{Section:4} with a summary,
and discussion of possible future extensions and applications.
Throughout we use geometric units, where the speed of light $c=1$ and Newton's constant $G=1$.

\section{{\bf The Schwarzschild Geometry}} 
\label{Section:1.5}
\begin{figure}[ht!]
\begin{center}
\includegraphics[width=13.00cm,clip=true]{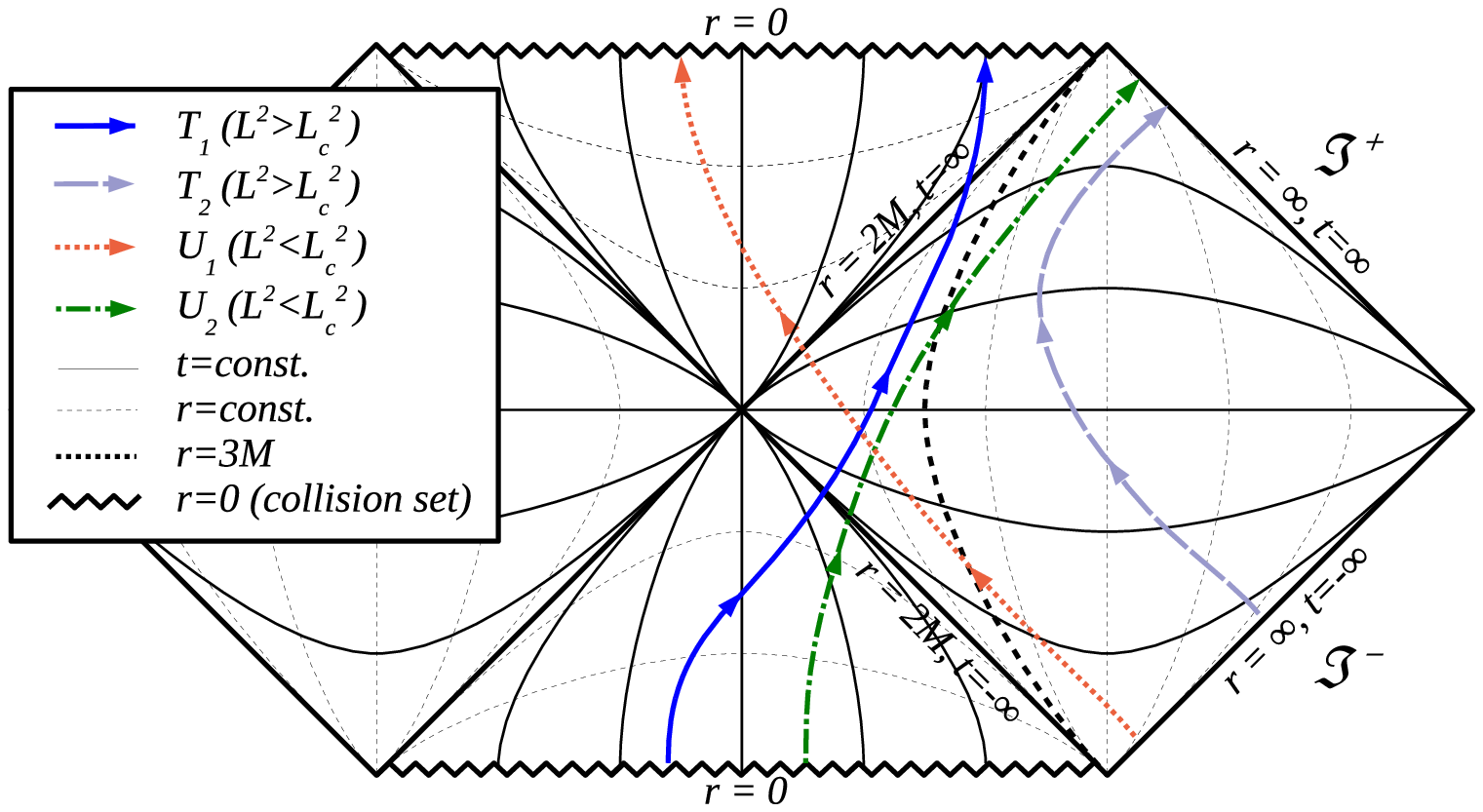}
\end{center}
\caption{A schematic Penrose diagram of the Schwarzschild solution, with representative
null geodesics from each of the invariant manifolds; see Figure ~\ref{Figure:McGehee} for
the corresponding diagram in McGehee coordinates. 
}
\label{Figure:Penrose}
\end{figure}

The Schwarzschild metric, describing a non-rotating black hole of mass $M$, has
the following line element in standard (spherical polar) Schwarzschild coordinates:
\begin{equation} 
ds^2 = g_{\alpha\beta} dx^\alpha dx^\beta = -(1-2M/r)dt^2 + (1-2M/r)^{-1}dr^2 + r^2(d\theta^2 + \sin^2\theta d\phi^2), 
\label{eq:1}
\end{equation}
where $g_{\alpha\beta}$ is the metric tensor, and we use the Einstein summation 
convention where repeat indices are summed over.
Figure ~\ref{Figure:Penrose} shows a Penrose diagram
of the Schwarzschild spacetime. For readers 
not familiar with this diagram, we will briefly review its
salient features. The Penrose diagram is a conformal compactification of the
maximal analytic extension of the Schwarzschild metric, designed to highlight
the causal structure of the spacetime. Here we project out the $(\theta,\phi$) coordinates,
so a point on the diagram represents a 
two-sphere of area $4\pi r^2$. Radial (zero angular momentum) null geodesics 
are straight lines angled at $\pm 45^\circ$ relative to the horizontal. Any causal 
curve that could be associated with a particle trajectory (whether timelike, null, geodesic or not) 
has a slope $\ge 45 ^\circ$ {\em everywhere} along its projected curve on the diagram. There are two
identical, causally disconnected, asymptotically flat regions reached as $r\rightarrow\infty$; without
loss of generality we focus on the right hand region on the diagram. Similarly, there
are two singularities in the geometry at $r=0$: the {\em white hole} singularity occurring to the 
past of any event in the spacetime, and the {\em black hole} singularity that is within the
causal future of any event. 

A null curved can be parameterized by an affine time $\sigma$, unique
up to a constant scaling and translation. 
Null curves that originate at $r=\infty$ (with $\sigma\rightarrow-\infty$)
come from a region
called {\em past null infinity} $\scrim$, and those that return to $r=\infty$ (as $\sigma\rightarrow+\infty$) 
end at {\em future null infinity} $\scrip$.
The event horizon of the black hole is formally defined
as the boundary of the causal past of $\scrip$, which on the Penrose diagram is the
line labeled $r=2M,t=\infty$. 
Null curves that come from the white hole ``begin'' at $r=0$ with finite $\sigma$,
and those that cross the event horizon reach $r=0$ in finite affine time.
Note that the Schwarzschild time coordinate $t$ is 
singular at $r=2M$. Further, note that $r$ and $t$ switch ``character'' at $r=2M$: outside,
$r$ is spacelike, $t$ timelike, and vice-versa inside. In the black hole, time as
measured by $r$ flows to smaller $r$ (the opposite in the white hole). Thus, the inevitability
of encountering the singular at $r=0$ for any observer crossing the horizon is evident
in the diagram: $r=0$ is no longer a ``place'' that can be avoided, rather it is a ``time''
that will happen for any such observer.
Finally, note that the compactification severely distorts the spacetime at the corner
points of the diagram---that $r=0$, $r=2M$ and $r=\infty$ touch in the diagram
is purely an artifact of the compactification.

\section{{\bf Geodesic Equations in the Schwarzschild Spacetime}}
\label{Section:2}
We begin by describing the geodesic equations for causal particles; later
we restrict to the null case.
The geodesic equation for a particle $P$ in parametric form $x^\alpha(\sigma)$ is
\begin{equation}\label{geod_eqn}
\ddot{x}^\alpha + \Gamma^\alpha_{\beta\gamma} \dot{x}^\beta \dot{x}^\gamma = 0 
\end{equation}
where the over-dot $\dot{} \equiv d/d\sigma$ and $\Gamma^\alpha_{\beta\gamma}$ is the metric connection.
This is formally a set of second order ordinary differential equations, though
due to the symmetries of the Schwarzschild metric, and the normalization
condition that $g_{\alpha\beta} \dot{x}^\alpha \dot{x}^\beta$ equals $0$ ($-1$)
for null (timelike) geodesics, one obtains the following first integrals
of motion of (\ref{geod_eqn}) in Schwarzschild coordinates (\ref{eq:1}):
\begin{eqnarray}
&& \dot{t} = {\tilde E} (1-2M/r)^{-1} \nonumber \\
&& r^4 {\dot{r}}^2 = {\tilde E}^2r^4-(r^2-2Mr)(\mu^2r^2+K) \nonumber \\
&& r^4{\dot{\theta}}^2 = K- L^2\sin^{-2}\theta \nonumber \\
&& \dot{\phi}  = L(r^2 \sin^2\theta)^{-1}, \label{geod_eqn_sch}
\end{eqnarray}
where $\mu^2$ is the rest mass of $P$, $\tilde E$ its energy, $L$ its angular
momentum about the axis $\sin \theta = 0$, and $K$ is Carter's constant of motion. 
Note that the second equation above for $\dot{r}$ can be written as a Newtonian central force problem
\begin{equation}\label{e_def1}
E=\frac{1}{2}\dot{r}^2 + V(r),
\end{equation}
with equation of motion
\begin{equation}
\ddot{r} = - dV/dr,
\label{eq:6.5B}
\end{equation}
where $E=\tilde E^2/2$ and the effective potential $V(r)$ is given by
\begin{equation}
V(r) = \frac{1}{2}\left(1-\frac{2M}{r}\right)\left(\mu^2+\frac{K}{r^2}\right).
\label{eq:6V}
\end{equation}

\noindent{\em Kerr-Schild coordinates}
\medskip

The Schwarzschild metric in Cartesian Kerr-Schild coordinates $(\lambda, x, y, z)$ is
\begin{equation}
ds^2 = -d\lambda^2 + dx^2 + dy^2 + dz^2 + 2Mr^{-3}(xdx + ydy + zdz+ r d\lambda)^2,
\label{eq:4}
\end{equation}
related to Schwarzschild coordinates by the set of transformations
\begin{equation}
x = r \sin\theta\cos\phi,   \hspace{1cm}  y =  r \sin\theta\sin\phi, \hspace{1cm}   z = r \cos\theta 
\label{eq:2}
\end{equation}
and 
\begin{equation}
d\lambda = dt +\frac{2M}{r-2M} dr.
\end{equation}
One advantage of Kerr-Schild coordinates over Schwarzschild coordinates, as is evident from (\ref{eq:4}), 
is that they are regular across the event horizon at $r = 2M$\footnote{Technically, (\ref{eq:4}) is
only regular across the black hole event horizon, and not the Cauchy horizon at $r=2M$ of
the white hole. The reason is $\lambda$ has been chosen so that along an $r={\rm const.}$
surface, $\lambda$ coincides with the retarded time $v$ of an {\em ingoing} radial
null curve coming from $\scrim$, and $v\rightarrow-\infty$ as the white hole horizon is approached
on the Penrose diagram. These coordinates are thus also referred to as
ingoing Eddington-Finkelstein coordinates. If instead we had chosen a coordinate 
$\bar{\lambda} = dt - 2M/(r-2M)dr$, then $\bar{\lambda}$ would coincide with advanced time $u$ of an 
outgoing radial photon along an $r={\rm const.}$ surface (outgoing Eddington-Finkelstein coordinates), 
and the metric
would be regular at the white hole horizon, but singular at the black hole horizon for
similar reasons. However, at the end of the day one arrives at the {\em same} 
geodesic equation of interest (\ref{eq:6.5}) for the spatial Cartesian
coordinates as a function of {\em affine time} $[x(\sigma),y(\sigma),z(\sigma)]$,
which is well defined across both horizons.}.

The geodesic equations in Kerr-Schild coordinates can be written as~\cite{Marck}
\begin{eqnarray}
&& \dot{\lambda }= 2MK r ^{-3} ({\tilde E} - \dot{r})^{-1} + {\tilde E}  \label{eq:5} \\
\medskip\medskip
&& \ddot{x}_k = -3MK x_k r^{-5} - \mu^2 M  x_k r^{-3} , \label{eq:6} 
\end{eqnarray}
where for the sake of notation, $k = 1,2,3$, with $x_1 = x,  x_2 = y, x_3 = z$. 
Due to the spherical symmetry of the Schwarzschild spacetime, without loss of generality we can restrict 
attention to planar motion. For simplicity we choose $z=0$, corresponding to $\theta=\pi/2$, for
which $K=L^2$ (\ref{geod_eqn_sch}). In the expressions below we will
replace $K$ with $L^2$, though one can consider them to be valid
for motion in any plane if $L$ is re-interpreted as the angular momentum
relative to an axis orthogonal to the orbital plane.
We will also now limit the discussion to
null geodesics, for which $\mu^2=0$, and only focus on the 
coordinate flow (\ref{eq:6}) of the geodesics:
\begin{equation}
\ddot{\bf{x}}= -3 M L^2 {\bf{x}} r^{-5} , 
\label{eq:6.5}
\end{equation} 
where ${\bf{x}} = (x_1, x_2) \in R^2$, and $r=|{\bf{x}}|=\sqrt{x_1^2+x_2^2}$. 
\medskip

The corresponding Hamiltonian for this system of equations is
\begin{equation}
E = H({\bf x}, {\bf{\dot x}}) = {1\over 2} |{\bf{\dot x}}|^2 - L^2 M|{\bf x}|^{- 3},
\label{eq:6H}
\end{equation}
where $E=\tilde{E}^2/2$ is a positive constant of motion for any particle trajectory.
\medskip

\noindent
It is noted that for each value $h \geq 0$ of the energy $E$,  the motion of $P$ of mass zero 
lies on the three-dimensional energy surface 
\begin{equation}
\Sigma = \{ ({ \bf x}, {\bf \dot x} ) \in R^4| H({\bf x}) =  h \geq 0 \}.
\label{eq:6EE} 
\end{equation}
\medskip
\medskip

To facilitate direct application of prior work on Newtonian central force
problems of the form (\ref{eq:6.5}), it is useful to perform the following
{\em dimensionful} rescaling of the affine parameter $\sigma$ to $\xi$ 
\begin{equation}\label{scaling}
\sigma=\frac{\xi}{\sqrt{L^2 M}}.
\end{equation}
This transforms (\ref{eq:6.5}) to
\begin{equation}
\frac{d^2 {\bf{x}}}{d\xi^2} = -3 {\bf{x}} r^{-5} , 
\label{eq:7}
\end{equation}
In the following section we consider the solutions ${\bf{x}}(\xi) = (x_1(\xi),x_2(\xi))$ of (\ref{eq:7}).

\section{{\bf Transformation to McGehee Coordinates and Blowup of Collision}}
\label{Section:3}

We consider general central force fields with potential
\begin{equation}
U({\bf x}) = -|{\bf x}|^{-\alpha},
\end{equation}
$\alpha >0$, later restricting to the case $\alpha=3$ of (\ref{eq:7}). 
The system of differential equation describing the motion of a single particle with this potential is given by
\begin{equation}
\ddot{\bf x} = -\partial_{\bf x} U({\bf x}) = -\alpha|{\bf x}|^{-\alpha-2}{\bf x}.
\label{eq:8}
\end{equation}
It is convenient to write this as the first order system 
\begin{eqnarray}
&& \dot{\bf x} = {\bf y}, \nonumber \\
&& \dot{\bf y} = -\alpha|{\bf x}|^{-\alpha-2}{\bf x}.
\label{eq:9}
\end{eqnarray}
This is a Hamiltonian system with Hamiltonian function
\begin{equation}
H({\bf x}, {\bf y}) = {1\over 2} |{\bf y}|^2 - |{\bf x}|^{- \alpha},
\label{eq:10}
\end{equation}
which is the total energy of the particle and is conserved along solutions of (\ref{eq:9}), where
$({\bf{x}}, {\bf{y}}) \in R^4$. 

The general description of the orbit structure for (\ref{eq:9}) was carried out
in \cite{McGehee}, with a special emphasis on the
motion of the particle near collision. 
The general approach taken is to find a change of coordinates which have the effect of blowing up the collision, corresponding to 
${\bf x} = {\bf 0}$, into an invariant manifold with its own flow. When this is done, the dynamics of the particle near collision can be completely
understood and solutions tending towards collision are asymptotic to this manifold. This change of coordinates also gives the global flow for the differential equations.   

Set $ {\bf X} = ({\bf x}, {\bf y})$, and  consider a solution ${\bf X} (\xi) = ({\bf x}(\xi), {\bf y}(\xi)) $ 
for (\ref{eq:9}) with an initial condition ${\bf X}(0)$. 
The standard existence and uniqueness theorems of 
differential equations guarantee that ${\bf X} (\xi)$ can be uniquely determined and defined over a maximal interval $(\xi^{-}, \xi^{+})$, where 
$-\infty \leq \xi^{-} < 0 < \xi^{+} \leq + \infty$.    

\medskip
\noindent
{\em Definition \hspace{.4cm} }  If $\xi^{+}< \infty$, then  ${\bf X} (\xi)$ {\em ends in a singularity} at $\xi^{+}$. If $\xi^{-}\geq -\infty$ then  ${\bf X} (\xi)$ {\em 
begins in a singularity} at $\xi^{-}$. In either case, $\xi^* = \xi^{+}$ or $\xi^{-}$ is said to be a {\em singularity} of the solution  ${\bf X} (\xi)$.     
\medskip

\noindent
The following result is proven in \cite{McGehee}:

\medskip
\noindent
Let  ${\bf X} (\xi)$ be a solution of (\ref{eq:9}) with a singularity at $\xi^*$. Then this singularity is due to collision. 
That is, ${\bf x} (\xi) \rightarrow 0$ as $z \rightarrow \xi^*$.
\medskip

It follows from (\ref{eq:10}) that for a collision solution, $|{\bf y}| \rightarrow \infty$ as $ z \rightarrow \xi^*$.
Hence, for (\ref{eq:9}) the only solutions that are singular are collision solutions, either ending or beginning in collision. 
There are several methods available to study collision, the dynamics of solutions near it and to understand if a collision solution can be
extended through the collision state in a smooth fashion. 
Here we present a brief summary of blowing up collision for (\ref{eq:9}) for arbitrary $\alpha > 0$, and then apply 
that to understand the global flow for the case of $\alpha =3$. The details are in \cite{McGehee}.

We set $\beta ={ \alpha\over2}$, $\gamma = (1 + \beta)^{-1}$ for sake of notation. Also, it is convenient to use 
complex coordinates and identify the real plane $R^2$ with the complex plane
$C^1$. Then, we can consider  $\bf x$ to be a vector in $R^2$ 
or a complex number in $C^1$. 
The {\em McGehee coordinates} are given by the transformation $T$  of $(x_1, x_2, y_1, y_2)$ to 
$({\tilde r} > 0, \tilde{\theta}, w, v)$, 
\begin{eqnarray}
\label{eq:11}
&& {\bf x} = {\tilde r} ^{\gamma}e^{i\tilde{\theta}}  \\
&& {\bf y} = {\tilde r}^{-\beta \gamma} (v + i w) e^{i \tilde{\theta}}
\nonumber
\end{eqnarray}
and a transformation of the affine variable $\xi$,
\begin{equation}
d\xi = {\tilde r} d\tau. 
\label{eq:12}
\end{equation} 
The system (\ref{eq:9}) is transformed into
\begin{eqnarray}
&& {\tilde r }' = (\beta+1) {\tilde r}v, \nonumber\\
&& \tilde{\theta}' = w, \label{eq:13} \\
&& w' = (\beta - 1)wv,
\nonumber \\
&& v' = w^2 + \beta (v^2 -2), \nonumber
\end{eqnarray} 
where prime denotes differentiation wrt $\tau$.  
In complex notation, the angular momentum for (\ref{eq:9}) is given by 
\begin{equation}
\Omega({\bf x}, {\bf y}) = \Im (\bar{{\bf x}} {\bf y}). 
\label{eq:momentum}
\end{equation}
\medskip
\noindent
We fix the energy $H$ to the constant value $h$ and $\Omega$ to the constant value $c$ (not to be confused with the speed of light). $T$ transforms $H = h$ and $\Omega = c$ into
\begin{eqnarray}
&&  w^2 + v^2 - 2 = 2h{\tilde r}^{\alpha \gamma} 
\label{eq:14} \\
&& {\tilde r}^{(1-\beta)\gamma} w = c. 
\label{eq:15}
\end{eqnarray}
We define the constant energy manifold,
\begin{equation}
{\bf M}(h) = \{ ({\tilde r},\tilde{\theta},w,v) \in R^4 | {\tilde r} \geq 0, H = h  \}.
\label{eq:16}
\end{equation}
We define the collision set corresponding to collisions for System (\ref{eq:9}) as
\begin{equation}
{\bf N} = \{ ({\tilde r},\tilde{\theta},w,v) \in {\bf M}(h)| {\tilde r}=0 \}.
\label{eq:17}
\end{equation}
On account of (\ref{eq:14}), (\ref{eq:15}) and (\ref{eq:17}) can be written as
\begin{eqnarray}
{\bf N} &=& \{ ({\tilde r},\tilde{\theta},w,v) \in {\bf M}(h)| {\tilde r}=0, v^2=2, w=0\} \ \ \ \beta>1,\\
{\bf N} &=& \{ ({\tilde r},\tilde{\theta},w,v) \in {\bf M}(h)| {\tilde r}=0, v^2+c^2=2, w=c\} \ \ \ \beta=1,\\
{\bf N} &=& \{ ({\tilde r},\tilde{\theta},w,v) \in {\bf M}(h)| {\tilde r}=0, w^2+v^2=2, c=0\} \ \ \ 0<\beta<1.
\label{eq:18}
\end{eqnarray}
As is proven in \cite{McGehee},
\medskip

\noindent 
{\bf N} is an invariant manifold for the vector field defined by System (\ref{eq:13}). Collision orbits approach {\bf N}
asymptotically as  $\tau \rightarrow \pm \infty.$
\medskip
\medskip

\noindent
{\em Definition}  \hspace{.5cm} ${\bf N}$ is called a {\em blow up} of the collision ${\tilde r}=0 $ on ${\bf M}(h)$.
\medskip

\noindent
It is also shown in \cite{McGehee} that System (\ref{eq:13}) reduces to the system consisting of the last two equations of (\ref{eq:13}),
\begin{eqnarray}
&& w' = (\beta -1) wv, \label{eq:19} \\
&& v' = w^2 + \beta(v^2 -2). \nonumber
\end{eqnarray}
This system has the integral
\begin{equation}
\Lambda(w,v) = |w|^\alpha |v^2 + w^2 -2|^{1-\beta}. 
\end{equation}

\noindent
In summary, to understand the flow of (\ref{eq:9}) on  ${\bf M}(h)$, we
can use System (\ref{eq:19}), where the collision manifold is given by  ${\bf N}$. 

\section{{\bf McGehee Flow of Schwarzschild Null Geodesics}}
\label{Section:3.5}
\begin{figure}[ht!]
\begin{center}
\includegraphics[width=13.00cm,clip=true]{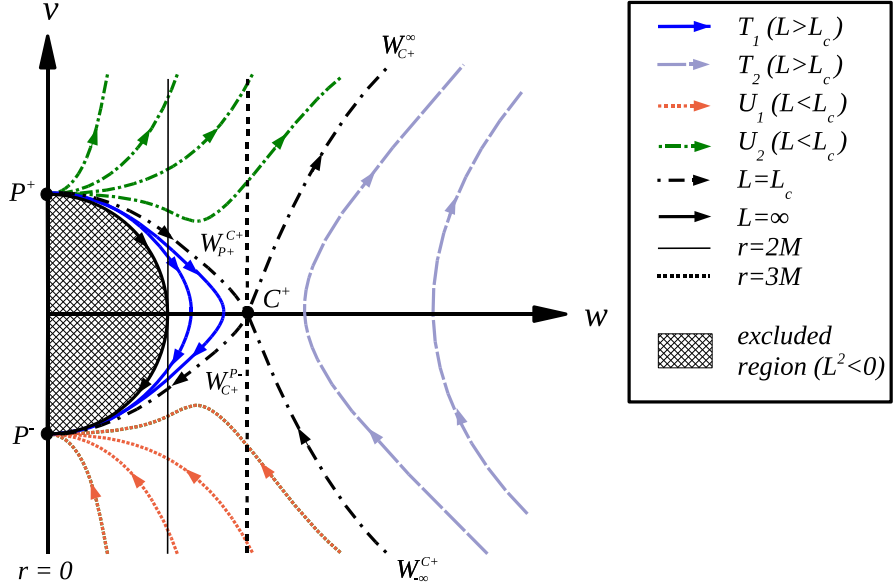}
\end{center}
\caption{The Schwarzschild null geodesic flow in McGehee Coordinates; see Figure ~\ref{Figure:Penrose} for the corresponding Penrose diagram. The point $P^-$ ($P^+$) here corresponds to the black hole (white hole) singularity at $r=0$.
For clarity we only show the half plane $w>0$, corresponding to positive
angular momentum $L$ orbits; the flow in $w<0$ for negative angular
momentum orbits looks identical to that of $w>0$, but reflected about
the $v$ axis.}
\label{Figure:McGehee}
\end{figure}

We now consider the case of interest given by the Schwarzschild problem that we showed reduced to System (\ref{eq:7}).
This corresponds to $\alpha=3$, $\beta = 3/2$, $\gamma = 2/5$.  
The flow for System (\ref{eq:19}) is depicted in Figure \ref{Figure:McGehee}.
Incidentally, this figure captures the qualitative behavior for all values of $\beta >
1$. It turns out that the flows for $\beta \leq 1$ are drastically different, in particular the Kepler problem 
corresponds to $\beta = 1/2$.
Over the next several paragraphs we will describe the flow in some detail,
though first it would be helpful
to identify a few relations between the  coordinates $(w,v)$ 
and physical coordinates, as well as establish the map between
the constants of motion $(h,c)$ of solutions to System (\ref{eq:9}) and the 
physically relevant constants $L,\tilde{E}$ and $M$. 

\medskip
\noindent{\em Relationships between constants and coordinates}
\medskip

\noindent 
Using (\ref{eq:11}) and (\ref{eq:15}), and the
angular momentum of the geodesic in the physical picture
$L=x\dot{y} - y\dot{x}$, one finds that
the angular momentum $c$ of the System (\ref{eq:9}) evaluates to
\begin{equation}\label{cdef}
c=\pm \frac{1}{\sqrt{M}}.
\end{equation}
This rather unusual relationship is due to the scaling (\ref{scaling}) between
the time parameters of the two descriptions. Likewise, it is straightforward
to show that the energy $h$ of the System (\ref{eq:9}) relates to 
physical constants via
\begin{equation}\label{hdef}
h=\frac{\tilde{E}^2}{2 L^2 M}.
\end{equation}
From (\ref{eq:11}), the 
relation between areal radius $r$ and $\tilde{r}$ for $\alpha=3$ is
\begin{equation}
r=\tilde{r}^{2/5},
\end{equation}
and this together with (\ref{cdef}) and (\ref{eq:15}) gives
\begin{equation}
w=\pm\sqrt{\frac{r}{M}},
\end{equation}
where the plus (minus) sign corresponds to positive (negative) physical
angular momentum $L$. Thus vertical lines of constant $w$ on the McGehee
diagram correspond to surfaces of constant areal radius $r$.
Putting all the above together, with (\ref{eq:14}), one finds
that all trajectories projected onto the $(w,v)$ plane are
characterized by the following polynomial with one free parameter $Q\ge 0$:
\begin{eqnarray}
v^2&=&2-w^2+Q w^6, \\
Q&\equiv&\frac{\tilde{E}^2 M^2}{L^2}.
\end{eqnarray}
Finally, note from (\ref{eq:13}) that $v=0$ implies $\tilde{r}'=0$,
which translates to $\dot{r}=0$; thus a trajectory that crosses $v=0$ 
corresponds to a geodesic that has a turning point in its radial motion.

\medskip
\noindent{\em Particle Flow}
\medskip

\noindent 
Figure \ref{Figure:McGehee} is a projection of the full flow in $(\tilde{r},\tilde{\theta},w,v)$ coordinates
to the $(w,v)$ plane.
Though as seen in (\ref{eq:13}), the two differential equations for 
$(w',v')$ do not depend on $(\tilde{r},\tilde{\theta}) $ and can be solved separately as an 
independent system. Once we know $(w,v)$ then the coordinates $(\tilde{r}, \tilde{\theta})$ can be 
easily computed for each $(w,v)$ by (\ref{eq:13}). $(\tilde{\theta}$ and $\tilde{r})$ represent polar-like 
coordinates for $P$, while $(w,v)$ can be viewed as velocity-like coordinates. 
System (\ref{eq:13}) implies that $\tilde{\theta} (\tau)$ either increases, $w > 0$, or decreases $w < 0$. 
This is just a cycling motion about the origin $\tilde{r} = 0$ in position coordinates
$\tilde{r}, \tilde{\theta}$. While this cycling motion is occurring, $\tilde{r}$ increases, $v >0$, or 
decreases, $v <0$.  
This is analogous to the projection on the Penrose diagram
of the flow in $(r,t,\theta,\phi)$ to the (conformally compactified) $(r,t)$ plane, and
cyclic motion for $K=L^2$ geodesics corresponds to increasing $\phi$, $L>0$, or decreasing $\phi$, $L<0$.

The flow curves $(w(\tau), v(\tau))$ in Figure \ref{Figure:McGehee} can be viewed as 
invariant manifolds. They foliate the $(w,v)-$plane. 
From (\ref{eq:13}) one can see that the flow has critical points $C^{\pm}$ at $(w,v) = (\pm \sqrt{3} , 0)$, 
where $\tilde{r}=\sqrt{3}$, and $\tilde{\theta} = \pm\sqrt{3}\tau$ (to within a constant phase) 
respectively. These correspond to the unstable (hyperbolic) circular periodic orbits at $r=3M$. 
The periodic orbits $C^\pm$ exist for each $h >0$ on each energy 
surface ${\bf M}(h)$ .    

\medskip
\noindent{\em Excluded region}
\medskip

\noindent 
The flow on ${\bf M}(h)$
for (\ref{eq:19}) projects into the set $\{w^2 + v^2 \leq 2\}$ for each $h < 0$, to $\{w^2 + v^2 = 2\}$ for $h=0$,  
and to $\{w^2 + v^2 >  2 \}$ for $h > 2$.  
Hence, since $P$ moves on $\Sigma$, defined by (\ref{eq:6EE}), where $h \geq 0$, then 
we need not consider the interior of the disc, $\{w^2 + v^2 < 2\}$. So, in Figure
\ref{Figure:McGehee} the region of interest are all points $\{w^2 + v^2 \geq 2\}$. 

\medskip
\noindent{\em The collision set}
\medskip

\noindent 
The collision set  ${\bf N}$ for $\alpha=3$ reduces to two critical points of the flow,
$p^+ = (0, \sqrt{2})$ and $p^- = (0, -\sqrt{2})$. These are unstable hyperbolic points. 
The flow tends to $p^+$ as $\tau \rightarrow -\infty$ and 
to $p^-$ as  $\tau \rightarrow
+ \infty$.  In the full phase space,
these critical points have a constant value of $\tilde{\theta}$, and $\tilde{r} = 0$. 
On the Penrose diagram, $p^-$ ($p^+$) corresponds to the black hole (white hole) singularity
at $r=0$. 

\medskip
\noindent{\em Invariant manifolds of the flow}
\medskip

\noindent 
There are invariant manifolds connecting  $C^{\pm}$ and $p^{\pm}$.    
Two flow from $p^+$ to $C^{\pm}$ and two flow from $C^{\pm}$ to $p^-$. These we label $W_{p^+}^{C^{\pm}}$  and  
$W_{C^{\pm}}^{p^-}$, respectively; trajectories within this flow have physical angular momentum-squared $L_c^2=27 \tilde{E}^2M^2$.
As $P$ moves along, e.g.  $W_{p^+}^{C^{+}}$, it can be viewed in position space as cycling about the origin while
moving outward toward the periodic orbit. The cycles converge to the periodic orbit asymptotically as
$\tau\rightarrow\infty$ for $W_{p^+}^{C^{+}}$, and $\tau\rightarrow-\infty$ for $W_{C^{+}}^{p^-}$. 

There are curves that leave $p^+$, move out to a maximum distance within the range $\sqrt{2} < w < \sqrt{3}$ ($2M<r<3M$) when $v=0$, then 
turn around and move to $p^-$. We label these manifolds $W_{p^+}^{p^-}$, and all of these
geodesics have $L^2>L_c^2$. The union of these manifolds fill a region labeled $T_1$ in Figure \ref{Figure:McGehee}.
In the limit $L\rightarrow \pm |L_c|$, the solutions asymptote to the union of $W_{p^+}^{C^{\pm}}$ and $W_{C^{\pm}}^{p^-}$. The $L^2\rightarrow \infty$ limit ($h\rightarrow 0$) are a rather interesting
class of geodesics, which we discuss in the following subsection.

Similarly, we have solutions that come in from negative infinity ($\scrim$) for $v < 0$,
reach a minimum distance from the black hole in the range $w>\sqrt{3}$ ($r>3M$) when $v=0$, then 
go back out to positive infinity ($\scrip$) with $v > 0$; they
lie on invariant manifolds we call $W_{-\infty}^{+\infty}$, and are labeled $T_2$
in Figure \ref{Figure:McGehee}.
As with $T_1$, all these geodesics have angular momentum squared $L^2 > L_c^2$.
In the limit $L \rightarrow \pm |L_c|$, they
asymptote to the union of the manifolds $W_{-\infty}^{C^{\pm}}$ and $W_{C^{\pm}}^{\infty}$, the
manifolds that connect $\scrim$ to $C^\pm$ and $C^\pm$ to $\scrip$ respectively.
The solutions on  $W_{-\infty}^{C^{\pm}}$ asymptotically ($\tau\rightarrow\infty$) approach the circular orbit as they 
spiral towards it from $\scrim$, and those on  $W_{C^{\pm}}^{\infty}$ (beginning at $\tau=-\infty$) asymptotically spiral away from the 
circular orbit to $\scrip$.

The remaining two regions of the flow we call $U_1$ and $U_2$. $U_1$ consists of the 
manifolds $W_{-\infty}^{p-}$, which spiral in from $\scrim$ into the black hole.
They do not encounter a turning point in $r$. These geodesics all have
$L^2<L_c^2$, and in the limit $L \rightarrow \pm |L_c|$ asymptotically
approach the union of manifolds $W_{-\infty}^{C^{\pm}}$ and $W_{C^\pm}^{p-}$.
Similarly, $U_1$ consists of the
manifolds $W_{p+}^{\infty}$, which spiral out from the white hole to $\scrim$,
do not have a turning point in $r$, have $L^2<L_c^2$, and 
in the limit $L \rightarrow \pm |L_c|$ asymptote
to the union of $W_{p+}^{C^{\pm}}$ and $W_{C^\pm}^{\infty}$.

\medskip
\noindent{\em The $L^2 \rightarrow\infty$ ($h=0$) limit}
\medskip

\noindent 
In region $T_1$, geodesics with $h\rightarrow 0$ correspond to the 
limit $\ell^2\equiv L^2/\tilde{E}^2\rightarrow\infty$.
Fixing the physical energy $\tilde{E}$ to be finite,
these geodesics have several curious properties\footnote{in contrast to the $\ell^2\rightarrow\infty$
limit region of $T_2$, which simply correspond to geodesics that pass by the black hole
with an infinite impact parameter}, most easily deduced from the geodesic
equations in Schwarzschild coordinates (\ref{geod_eqn_sch}). First, for $r\le2M$,
\begin{equation}
\frac{dt}{dr} =\pm \frac{r^{3/2}}{\sqrt{r(2M-r)^2 + \ell^2(2M-r)}},
\end{equation}
where the $+$($-$) sign corresponds to the part of the trajectory in the
white (black) hole where $\dot{r}>0$ ($\dot{r}<0$).
Taking the limit $\ell\rightarrow\infty$, one gets that these geodesics projected onto the
$(r,t)$ plane in the Penrose diagram correspond to $t={\rm constant}$ lines (and
recall that inside the horizon $t$ is a spacelike coordinate); 
i.e. they emanate from the white hole singularity, turn around
at the intersection of the event and Cauchy horizons at $r=2M$, then continue to the black hole
singularity. Next, we calculate how much cycling motion in $\phi$ they execute.
From the geodesic equations
\begin{equation}
\frac{d\phi}{dr} = \pm \frac{\ell}{\sqrt{r^4 + r \ell^2(2M-r)}},
\end{equation}
again where $+$($-$) corresponds to the white (black) hole regions.
Taking the limit $\ell\rightarrow\infty$, and integrating 
from $r=0$ to $2M$ in the white hole and back from $r=2M$ to $0$ in the black hole
gives $|\Delta \Phi|=2\pi$ for the journey---these geodesics circle in azimuth 
exactly {\em once} going from the white to black hole singularity.
Finally, we compute the total affine time $\Delta\sigma=\int d\sigma$ from
\begin{equation}
\frac{d\sigma}{dr} = \pm \frac{r^{3/2}}{\tilde{E}\sqrt{r^3 + \ell^2(2M-r)}}.
\end{equation}
For finite $\tilde{E}$, in the limit $\ell\rightarrow\infty$,
$\Delta\sigma=0$! This rather bizarre result could be remedied by an
infinite rescaling of $\sigma\rightarrow\sigma(\ell_0/\ell)$, where $\ell_0$
is some finite constant with dimension of length to make the scaling dimensionless.
It is unclear what the physical significance such a rescaling is.

\medskip
\noindent{\em Topology of the flow}
\medskip

\noindent 

\begin{figure}[ht!]
\begin{center}
\includegraphics[width=13.00cm,clip=true]{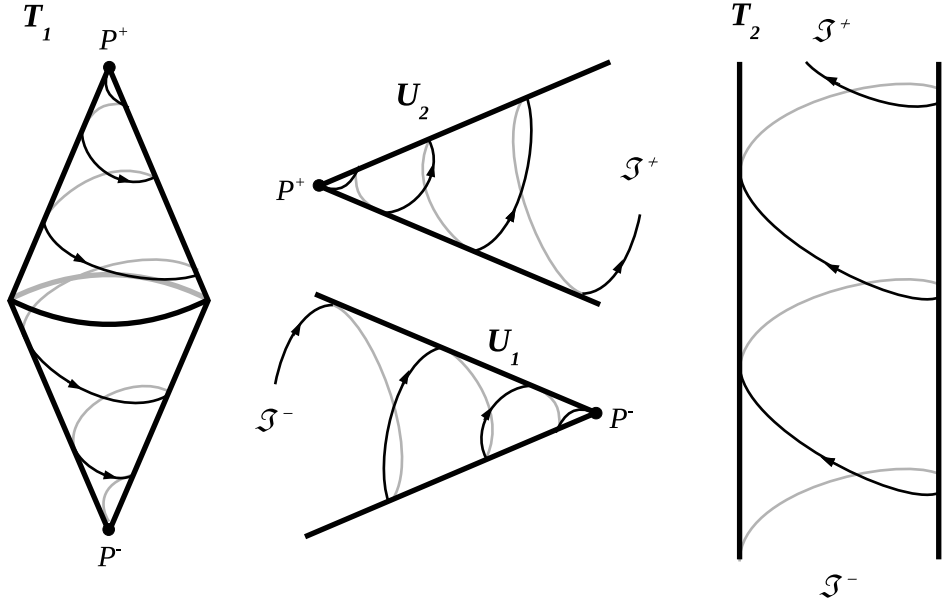}
\end{center}
\caption{Topology of the flow traced out by solutions within each of the
four invariant manifolds $T_1, T_2, U_1$ and $U_2$ (see Figure \ref{Figure:McGehee}).
Note that the sample trajectories are schematic only and for visual aid;
only those solutions that flow close to the critical points
$C^{\pm}$ trace out multiple cycles, and these are clustered
near $w=\sqrt{3}$ ($r=3M$). 
}
\label{Figure:Bridge}
\end{figure}

The solutions in the flow $T_1$ that spiral out from the white hole,
turn around at $\sqrt{2}\le w\le\sqrt{3}$, and return to the
black hole, trace out surfaces that are topologically equivalent to 
two cones smoothly joined together at $v=0$, with vertices anchored
in the critical points at $p^+,p-$; see Figure \ref{Figure:Bridge}.
The flows $U_1$ coming from $\scrim$ into the black hole topologically form an 
(infinite length) open cone, with the open end at $\scrim$, and the vertex
at $p^+$; similarly for the flows in $U_2$. The geodesics
in $T_2$ that come in from $\scrim$ and return to $\scrip$ lie
on families of surfaces that are topologically equivalent to
(infinite) cylinders.

\medskip
\noindent{\em Extension of Solutions Through Collision}
\medskip

Let ${\bf{x}}(\xi)$ be any solution of (\ref{eq:8}) which ends in collision at $\xi=0$. That is, ${\bf{x}}(\xi) \rightarrow 0$ as $\xi \rightarrow 0+$, where 
$\xi \rightarrow 0+$ means that $\xi$ approaches $0$ where $\xi > 0$. It is proven in \cite{McGehee} that  ${\bf{x}}(\xi)$ is {\em branch regularizable} if and only if 
$\gamma \in \Xi$, where we define the set $\Xi$ as follows: Let $m,n$ be relatively prime integers, then $\Xi = \{ m/n | n\ \hbox{odd}, n > m > 0 \}$. 
Recall that $\gamma = (1 + \beta)^{-1}, \alpha = 2\beta$. In our case, $\alpha = 3$ and $\gamma = 2/5$ belongs to the set $\Xi$.    
\medskip

\noindent
We define branch regularizable as follows:  A solution ${\bf{x}} = {\bf{W}}(\xi)$ of (\ref{eq:8}) which either begins or ends in collision at $\xi = \xi^*$
is branch regularizable at $\xi^*$ if it has a unique {\em branch extension} at $\xi^*$. A branch extension is defined by considering two solutions
${\bf{W_1}}(\xi)$
and ${\bf{W_2}}(\xi)$ of (\ref{eq:8}), where  ${\bf{W_1}}(\xi)$ ends in collision at $\xi^*$ and  ${\bf{W_2}}(\xi)$ begins in collision at $\xi^*$. Then,
${\bf{W_2}}(\xi)$ is a branch extension of ${\bf{W_1}}(\xi)$ if ${\bf{W_2}}(\xi)$ is a real analytic continuation of ${\bf{W_1}}(\xi)$ 
at $\xi=\xi^*$ for $\xi$ 
in a neighborhood of $\xi=\xi^*$. 
\medskip

\noindent
The fact that ${\bf{x}}(\xi)$ is branch regularizable implies that there is a way that 
solutions can smoothly and uniquely be extended through  $p^+$ in backwards time and through $ p^-$ in 
forwards time. For example,
let  ${\bf{Z}}(\xi) = ({\bf{x}}(\xi), {\bf{y}}(\xi))$ be a trajectory in phase space defined
for $\xi<0$ that collides with $p^-$ at $\xi=0$. Then there exists a 
unique extension of ${\bf{Z}}(\xi)$ for $\xi > 0$. This
extension is real analytic as a function of $\xi$ in a neighborhood of $\xi = 0$,  and corresponds to a 
smooth bounce of $P$ in $(x_1, x_2)$ space. In the spacetime picture, collision
with $p^-$ corresponds to the geodesic encountering the black hole singularity. The field
equations of general relativity do not describe how spacetime can be extended beyond
a singularity, and it is usually thought that a theory of quantum gravity is
required to ``resolve'' the singularity. Nevertheless, one way to map the branch regularized extension to 
geodesic motion ``through'' the singularity would be to identify the black hole
singularity with a white hole singularity of a second Schwarzschild solution of
identical mass\footnote{Identification with the white hole of the {\em same} solution is also a mathematical
possibility, though that would create closed timelike curves within the spacetime,
considered by some a class of ``pathology'' more severe than the black/white hole singularity.}. 

\section{\bf Conclusions}
\label{Section:4}
In this paper we have studied the relationship between the null geodesic structure of the Schwarzschild 
black hole solution, and the corresponding inverse-cubic Newtonian central force problem,
using the methods of McGehee. Both these problems have been well studied before, though what we
believe is novel in this paper is highlighting the {\em exact} correspondence between
the two descriptions, allowing insights from the dynamical systems approach to be
brought to the geodesic problem, and vice-versa. Indeed, in that regard it is rather amusing
to note that McGehee titled his paper ``Double Collisions for Classical Particle System with Nongravitational 
Interactions''. It is also rather interesting that what in the Newtonian picture may be regarded
as a clever ``trick'' using coordinate transformations to blow up the singular point of collision between
two particles, is in a sense the natural way to describe Schwarzschild spacetime. 
Another remark is that understanding the invariant manifolds and unstable hyperbolic points
allows standard techniques to be used to show that perturbations of the geodesic flow will
generically cause chaotic motion. For example, for this purpose,
solutions in $T_1$  may be regarded as forming a homoclinic loop if we identify $p^+$ and $p^-$. 
Perturbations will then generically break the homoclinic loop and cause chaotic motion by the 
Smale-Birkhoff theorem~\cite{Belbruno}.

Even though we restricted attention to null geodesics for simplicity, we expect that similar
mappings could be used for timelike particles, or more complicated geometries like Kerr. This would
be an interesting avenue for future work.

\section{\bf Acknowledgements}
We would like to thank I. Rodnianski and D. Spergel for helpful discussions. 
This work was supported by the Alfred P. Sloan Foundation (FP), NSF
grant PHY-0745779 (FP), and NASA/AISR grant NNX09AK61G (EB).

\medskip
\noindent

\end{document}